\documentclass[twocolumn,showpacs,preprintnumbers,amsmath,amssymb]{revtex4}
\usepackage{amsfonts}
\usepackage{amssymb}
\usepackage{latexsym}
\usepackage{graphicx}
\usepackage{psfig}
\usepackage{natbib}
\begin{document}

\preprint{M\'exico ICN-UNAM, \ April 2006}
\title{The Exotic Molecular Ion  $H_4^{3+}$ in a Strong Magnetic Field}
\author{H.Olivares Pil\'on}
\affiliation{Instituto de Ciencias Nucleares, UNAM, Apartado Postal 70-543, 04510 M\'exico}

\begin{abstract}
\centerline{\it Abstract} Using the variational method, a detailed
study of the lowest $m=0,-1$ electronic states of the exotic
molecular ion $H_4^{3+}$ in a strong magnetic field, in the linear
symmetric configuration parallel to the direction of the magnetic
field is carried out. A extended study of the $1\sigma_g$ ground
state (J.C.  L\'opez and A. Turbiner, {\it Phys.  Rev. \bf A 62},
022510, 2000) was performed obtaining that the potential energy
curve displays a sufficiently deep minimum for finite internuclear
distances, indicating the possible existence of the molecular ion
$H_4^{3+}$, for magnetic fields of strength $B\gtrsim
3\times10^{13}$\,G.  It is demonstrated that the excited
state $1\pi_u$ can exist for a magnetic field
$B=4\textrm{.}414\times10^{13}$\,G corresponding to the limit of
applicability of the non-relativistic theory. 
\end{abstract}

\maketitle

\section{Introduction} 

The neutron stars were predicted by L.~Landau~\cite{Landau:NS}
in 1932. This kind of stars are characterized by an enormous
magnetic field ($B \sim 10^{12}$\,G $ - 10^{13}$\,G). Since the
discovery of neutron stars in 1967 the behavior of the molecular
and atomic systems in such strong magnetic field has been 
object of intense investigations. The first studies were made in 
plasma physics and astrophysics by Kadomstev-Kudrayavstev ~\cite{Kadomtsev:1971} and Ruderman ~\cite{Ruderman:1971},
respectively. The results demonstrated that the binding energy
of a molecular system grows substantially with magnetic
field increase, while the size of the systems decreases. As 
a consequence of such a behavior, the formation of a molecular
chain aligned in the magnetic field direction arise.

Recent studies of the Coulomb systems in a strong magnetic
field have been encouraged by the observation of the 
neutron star 1E1207.4-5209 (\textit{Chandra} X-ray
observatory 2002). These observations displayed two
absorption features at 0.7\,keV and 1.4\,keV. The origin
of such lines could be associate to the presence of exotic
molecular systems in the atmosphere of the above mentioned star~\cite{Turbiner:2004b}. In particular, the studies
of one-electron molecular systems, traditional and exotic have been intensified, such as $H_2^{+}$, $H_3^{2+}$, $H_4^{3+}$,
$(HeH)^{2+}$, $He_2^{3+}$ y $Li_2^{5+}$ (see for example~\cite{Turbiner:2005}).

The focus of this article is a quantitative analysis of the
lowest electronic states of the molecular system $H_4^{3+}$
placed in a strong magnetic field. The variational method 
in the framework of non-relativistic theory is used, {\it i.e.} $B\leq 4\textrm{.}414\times10^{13}$\,G (Schwinger limit).
Specifically, an extension of the preliminary study
~\cite{LopezTur:2000} for the ground state is made, and the
possible existence of excited states is carried out.

Atomic units are used throughout ($\hbar=e=m_e=1$) and the energy
is expressed in Rydbergs (Ry). Spectroscopic convention is
used to denote the states. A number indicates the
principal quantum number, Greek letter $\sigma, \pi, \delta \ldots$ indicate the magnetic quantum number for $|m| = 0,1,2 \ldots$
respectively. The spatial parity is defined by $g$
(gerade) for states with even parity and $u$ (ungerade)
for odd parity states. Particularly, the ground state
is indicated as $1\sigma_g$.


\section{The Molecular Ion $H_4^{3+}$}

\begin{figure}[thb]
\begin{picture}(200,180)
\put(0,10){{\psfig{figure=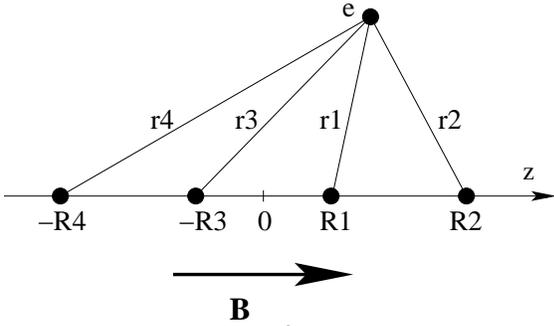,width=210pt}}}
\end{picture}
\vskip -20 pt 
\caption{The molecular ion $H_4^{3+}$ in a linear 
configuration placed in a constant and homogeneous
magnetic field directed according to $z$ axis,
$\mathbf{B} = (0,0,B)$. The protons positions are
indicated by the four points over $z$ axis.The position
of the electron is marked by a point out
of the axis.}
\label{El_ion}
\end{figure}

The exotic molecular ion $H_4^{3+}$ is a system
constituted by four protons and one electron. For
the present work the protons are consider over a
straight line (linear configuration) parallel to
the lines of magnetic field $B$ (parallel 
configuration). The magnetic field $B$ is taken
as constant and homogeneous in $z$ direction.
The Born-Oppenheimer approximation of zero order is
assumed (protons are consider infinitely massive).
The positions of the four protons over $z$ axis
are denoted as R$_1$, R$_2$, $-{\rm R}_3$ and
$-{\rm R}_4$. We will suppose also that 
R$_1=\,\,$R$_3$ and R$_2=\,\,$R$_4$
(symmetric configuration (see Fig.\ref{El_ion})).

In order to describe the magnetic field $\mathbf{B}=(0,0,B)$,
the vector potential in the symmetric gauge $\mathbf{A}=\frac{B}{2}(-y,x,0)$ is assumed.
The Hamiltonian that describes the molecular ion $H_4^{3+}$
(in cylindrical coordinates) is

\begin{equation}
\label{Hua}
\hat{\mathcal {H}}= -\mathbf{\Delta} +  \mathit{\hat{l}} _zB +
\frac{B^2\rho^2}{4}  + V(\rho,z) \,,
\end{equation}
where $\rho^2 = x^2 + y^2$ is the distance from
the electron to the $z$ axis. $\it{\hat{l}}_z$ is the $z$
component of the angular moment (which is a integral of
motion with eigenvalue $m$ -–magnetic quantum number).
$V(\rho,z)$ is the Coulomb potential of
the interaction between the electron and each charged center,
plus the classic repulsion term between the charged centers.
The potential is given by 
\begin{eqnarray}
\label{V}
  V(\rho,z) & = & -\frac{2}{r_1}- \frac{2}{r_2}-
  \frac{2}{r_3}- \frac{2}{r_4}  \\
      & & {} +\frac{2}{R_2-R_1}+
  \frac{2}{R_2+R_3}+
  \frac{2}{R_2+R_4} \nonumber  \\
      & & {} + \frac{2}{R_1+R_3}+
  \frac{2}{R_1+R_4}+
  \frac{2}{R_4-R_3}\, , \nonumber
  \end{eqnarray}
where $r_{i}=\sqrt{\rho^2+(z-z_i)^2}$ is the distances between the
electron and the corresponding proton ($z_i$ are the coordinates
of each proton over the $z$ axis).

In the Hamiltonian (\ref{Hua}) the spin contribution is neglected
because it only changes a reference level for the total energy.


\section{The Trial Functions}

In order to study the system $H_4^{3+}$ the variational
method is used. For the choosing of the trial functions, the fitting criteria used by Turbiner-L\'opez was followed (see for example
~\cite{Turbiner:2005}). These criteria have proved to give 
much precise results, qualitatively and quantitatively,
for different Coulomb systems in strong magnetic fields.
Basically these are:

\begin{itemize}
\item[1.-]The potential $V_t=\frac{\Delta\psi_t}{\psi_t}$, for
which the trial function $\psi_t$ is an exact eigenfunction of
$\mathcal{\hat H}_t = -\Delta + V_t$, with zero energy, should
reproduce the original potential near the Coulomb singularities
and the harmonic oscillator behavior at large distances.
\item[2.-]The symmetries of the system must be included.
\item[3.-]For the ground state, the trial function should not
vanish inside the domain where the problem is defined(Theorem 
of Perron).
\end{itemize}

In addition, since the charged centers (protons) are identical,
the trial function is chosen symmetric with respect to the
interchange of these, {\it i.e.}, invariant under
transformations $P_{ij}(r_i \leftrightarrow r_j),
\, i, j=1\ldots 4$.

The trial function is a linear combination of terms that
fulfill the requirements above mentioned. The general form
of the {\it ansatz} is
\begin{equation}
\Psi_{m} \equiv \,\mathcal{U}_m \,\Psi_{_0}\,,
\label{ET}
\end{equation}
where
\begin{subequations}
\begin{eqnarray}
\label{FPG}
\Psi_{_0} &=& \Big(\sum_{perm\{\alpha_1,..,\alpha_4\}}
e^{-\alpha_1 r_1-\alpha_2 r_2 -\alpha_3 r_3 -\alpha_4 r_4}\Big) e^{
  - \beta B   \frac{\rho^2}{4}} , \,\,\,\,\,\,\,\,\,\,
\\
\mathcal{U}_m &=& e^{\mathit{i}m\phi}\rho^{|m|} \,.
\label{FPGb}
\end{eqnarray}
\end{subequations}
Here $\Psi_{_0}$ corresponds to the general trial function
for the ground state ($m=0$). $\alpha_{1,2,3,4}$ and
$\beta$ are variational parameters. The sum is over all
permutations of the parameters $\alpha_{1,2,3,4}$.
Excitations for different values of the magnetic
quantum number $m$ are included with the $\mathcal{U}_m$
term. The trial functions $\Psi_{m}$ given by Eq.~(\ref{ET})
corresponding to different values of $m$ are orthogonal
and result suitable to describe the lowest states inside
each family of own states with quantum number $m$ defined.


\subsection{Gauge Rotation}

Considering the trial function $\Psi_{m}$ given by Eq.~(\ref{ET})
and the Hamiltonian $\mathcal{\hat{H}}$ (see Eq.~(\ref{Hua})),
the total energy (for a state with a quantum magnetic number $m$) is

\begin{equation}
   E_{T_{var}} = \begin{array}{c} min\\ \{\alpha's,\beta's \} \end{array}
   \frac{\displaystyle \int \rho^{2|m|}\, \Psi_{_{0}}
   {\hat{\mathit{h}}}_m \, \Psi_{_{0}} d\mathbf{r}}{\displaystyle \int
    \rho^{2|m|}\,  \Psi_{_{0}} \Psi_{_{0}} d\mathbf{r} } \,,
 \label{EVarn}
 \end{equation}
where
\begin{equation}
\label{hm}
{\hat{\mathit h}_m} \equiv \mathcal{U}_m^{-1} \,\hat{\mathcal
  {H}}\, \mathcal{U}_m =
 \mathbf{\hat{p}}_m^2
 + B m +\frac{B^2}{4}
\rho^2 + V(\rho,z)  \,,
\end{equation}
is the "gauge rotated Hamiltonian", and $\, \mathbf{\hat p}_m \equiv \mathcal{U}_m^{-1} \mathbf{\hat{p}}\,\, \mathcal{U}_m $ is the " covariant moment" . From here on, $E_{T_{var}}$ will denote simply as $E_{T}$. This formulation is very suitable since each Hamiltonian ${\hat{\mathit h}_m}$ (for $m$ constant) describes a family of eigenstates with the quantum number $m$.

Moreover, the ground state and the excited states are described by
trial functions which have the same structure (given by the
function~(\ref{FPG})). The only difference appears as a
weight factor $\rho^{2|m|}$.

The potential associated to an arbitrary trial function 
$\phi_{_{0}}$ is  now determined by 
\begin{equation}
V_0^{(m)}=\frac{{\mathbf{\hat p}_m }^2 \Phi_{_{0}}  }{\Phi_{_{0}} } \equiv
\frac{\tilde\Delta\Phi_{_{0}}}{\Phi_{_{0}}}\,,
\end{equation}
where the Laplacian covariant $\tilde\Delta$  is defined as
\begin{equation}
\label{LapCov}
\tilde\Delta=  \Delta +2 \displaystyle |m|\left(\frac{1}{\rho}
\frac{\partial}{\partial \rho} + \frac{\mathit{i}}{ \rho^2}
\frac{\partial}{\partial \phi}\right)\,.
\end{equation}
Hence, the general form of the potential associated with each
trial function term $\Psi_{_{0}}$, (Eq.~(\ref{FPG})) is~\footnote{This potential corresponds to the first term of the
trial function (Eq.~(\ref{FPG})) with parameters $\alpha$ without
permute.}
\begin{eqnarray}
\label{Vom}
V_{0}^{(m)} &=& \Big\{\beta^{} B\rho^2-2
\Big\}\sum_{i=1}^{4}\frac{\alpha_i^{}}{r_i}
+\sum_{i,k = 1}^4\alpha_i^{}\alpha_k^{}(\hat{\mathbf
  n}_i\cdot\hat{\mathbf n}_k)\,\,\,\,\,\,\,\,\,\,\,\,\,\\
&-&\beta B +
\frac{(\beta^{})^2 B^2}{4}\rho^2 -
2|m|\sum_{i=1}^4\frac{\alpha_i^{}}{r_i}
-|m|\beta^{} B \nonumber\,,
\end{eqnarray}
where
\begin{displaymath}
\hat{\mathbf n}_i\cdot\hat{\mathbf n}_k=
\frac{1}{r_ir_k}\{\rho^2+(z-z_i)(z-z_k)\}\,, \nonumber
\end{displaymath}
and $\hat{\mathbf n}_i\equiv ({{\mathbf r} - {\mathbf r}_i})/{|{\mathbf
r} - {\mathbf r}_i|}$ is a unitary vector oriented according
to the line which connects the $i$th position charged center,
with the position of electron ${\mathbf r}$. The potential
(\ref{Vom}) exactly reproduces the original potentials of
Coulomb for $\alpha_1=\alpha_2=\alpha_3=\alpha_4=1$,
and of harmonic oscillator for $\beta=1$.


\section{Trial Function Terms $\mathbf{\Psi}_{p} $}

The election of the most general trial function $\Psi_{_0}$
(Eq.~(\ref{FPG})) was in no way arbitrary. The $\alpha$
parameters may be interpreted as the effective charge of the
protons and $\beta$ parameter as the effective coupling of the
electron with magnetic field $B$. With this characterization of the variational parameters, it is possible obtain different trial functions 
each one with certain physical interpretation. 

Let us consider for example the case in which the interaction of
the electron with each one of the protons is {\it coherent}. That is,
the electron interacts in the same way with each proton. To
describe such physical situation, it is enough to take
$\alpha_1=\alpha_2=\alpha_3=\alpha_4\equiv\alpha_1 $ in the
Eq.~(\ref{FPG}). Defining this function as $\psi_{_0}^{(1)}$ the
following is obtained

 \begin{displaymath}
 \psi_{_0}^{(1)} =
e^{-\alpha_{1}(r_1+r_2+r_3+r_4)}\,\,\,e^{-\beta_{1}\frac{B}{4}\rho^2}\,.
 \end{displaymath}
This function is the Heitler-London type function (used to describe
the system $H_2^+$ without magnetic field) multiplied by the
lowest Landau orbital. The results of the calculus verify (a
posteriori) that this function gives the major contribution in the
description of the system near its equilibrium position.

In order to describe the physical situation at large internuclear
distances, all $\alpha's$ parameters are equal to zero, except one,
which is denote as $\alpha_2$. The trial function is  

 \begin{eqnarray*}
  \psi_{_0}^{(2)}
  =\Big\{e^{-\alpha_{2}
  r_1}+e^{-\alpha_{2}r_2}+e^{-\alpha_{2}r_3}+e^{-\alpha_{2}r_4}
  \Big\}\,\,\,e^{-\beta_{2}\frac{B}{4}\rho^2}\,,
  \end{eqnarray*}
which corresponds to a Hund-Mulliken type function (used to
describe the system $H_2^+$ without magnetic field) multiplied
by the lowest Landau orbital. This function describes the
interaction of the electron with each of the protons
independently ({\it incoherent interaction}).

It is now clear that different degenerations from the general
function (Eq.~(\ref{FPG})) lead to different physical descriptions
of the system, and therefore to several trial functions. In this
case and in order not to discriminate between one and another of
the different descriptions, the linear combination of all the
possible particular cases was considered as trial function.
Explicitly, the  trial function used in the current study is
\begin{eqnarray}
\label{LFP}
\Psi_{t}&=&A_1\psi_{_0}^{(1)} + A_2\psi_{_0}^{(2)} +
A_3\psi_{_0}^{(3)} + A_4\psi_{_0}^{(4)} +  \\
&&  A_5\psi_{_0}^{(5)} + A_6\psi_{_0}^{(6)}
+A_7\psi_{_0}^{(7)} + A_8\psi_{_0}^{(8)} + \nonumber\\
&&  A_9\psi_{_0}^{(9)} + A_{10}\psi_{_0}^{(10)} +
A_{11}\psi_{_0}^{(11)}.\nonumber
\end{eqnarray}
All the terms included in the last equation admit a certain physical
interpretation. This are summarized in Table \ref{TIF} where we
also present the variational parameters number contained in each
one of them. A detailed analysis of each term appears in~\cite{Tesis}.

\begin{table}[thb]
\begin{center}
\begin{tabular}{| c | c | c |}
\hline
Function &N.V.P. &Physical Description \\
\hline \hline
$\psi_{_0}^{(1)}$ &2 &Coherent Interaction \\
\hline \hline
$\psi_{_0}^{(2)}$ &2 &Incoherent Interaction \\
\hline \hline
$\psi_{_0}^{(3)}$ &2 &$H_2^+ + 2p$ \\
\hline \hline
$\psi_{_0}^{(4)}$ &2 &$H_3^{2+} + p$ \\
\hline \hline
$\psi_{_0}^{(5)}$ &3 & `` $H_2^+ + H_2^+$ '' \\
\hline \hline
$\psi_{_0}^{(6)}$ &3 & ``$H_3^{2+} + H$'' \\
\hline \hline
$\psi_{_0}^{(7)}$ &4 & ``$H_2^{+} + H + H$''  \\
\hline \hline
$\psi_{_0}^{(8)}$ &3 &$p\, +$  ``$H_2^{+} + H$''  \\
\hline \hline
$\psi_{_0}^{(9)}$ &4 &$p\, +$  ``$H + H + H$'' \\
\hline \hline
$\psi_{_0}^{(10)} $ &3 &  $p + p\, +$  ``$H + H$'' \\
\hline \hline
$\psi_{_0}^{(11)} $ &5 & ``$H + H + H + H$'' \\
\hline
\end{tabular}
\end{center}
\caption{ The trial function terms $\Psi_t$ (Eq.~\ref{LFP}) along
with the number of variational parameters (N.V.P.) and its
physical interpretation (see text).} \label{TIF}
\end{table}

Another degeneration of the Eq.~(\ref{FPG}) happens when  
$\alpha$'s parameters are equal by pairs,
$\alpha_1=\alpha_2=\alpha$; $\alpha_3=\alpha_4=\alpha'$,
($\alpha\ne\alpha'$). The physical picture correspond to an electron
shared by two protons of different form. The situation may be
interpreted as an interaction $H_2^+ + H_2^+$  between two
Hydrogen molecular ions, even though of course we only have one
electron (these type of physical interpretations are denoted
between quotation marks on Table \ref{TIF}). In the same way, in
the most general case (all $\alpha$'s parameters are
different), the corresponding function is a sum of products of four
Hydrogen functions by one Landau function (see Eq.~(\ref{FPG})).
This situation describes a mixed state of four Hydrogen "atoms" 
sharing only one electron.

Along with the $A_{1,\ldots ,11}$ coefficients of the linear
combination (Eq.~(\ref{LFP})) and the positions $R_{1,\ldots,4}$
(see Fig.~\ref{El_ion}) there is a total of 48 variational
parameters. The normalization condition of the trial function
allows to keep one of the $A$ coefficients as constant. The symmetry restriction ($R_3=R_1, R_4=R_2$) fixes two parameters. Finally the
number of effective variational parameters is reduced to 45.


\section{Computer Technique}

All calculations were made in a DELL PC with two Xeon processors
of $2.8$ GHz each. For minimization calculations the minimization
routine MINUIT of the CERN-LIB library was used. Numerical
integration is carried out using the routine D01FCF from the NAG-LIB library. The relative precision in the numerical integration was of
$10^{-11}$. It is important to mention that once all variational
parameters are determined, the calculation of the variational
energy takes a few seconds.

\subsection{Minimization Strategy}

The variational energy calculation includes two technical aspects:
the numerical evaluation of integrals in two dimensions and 
minimization of energy with respect to the variational parameters.
The objective in the minimization process is to find a point in the
multidimensional parametric space that minimizes the total energy
of the system (Eq.~(\ref{EVarn})). Since we have 45 variational
parameters, the search of the minimum is a formidable, complicated
and tedious task. For this reason it is not a good idea to start
the procedure making a minimization in which all parameters vary
simultaneously. The strategy used consisted in making minimization
sequences in the following way

\begin{itemize}
\item[$\bullet$]Ansatz by Ansatz, knowing clearly that between the
eleven terms that appear in the Eq.~(\ref{LFP}), the most general
$\psi_{p_0}^{(11)}$ should give us the best (lowest) value in
total energy.
\item[$\bullet$]Combinations of two by two
As\"antze, taking $\psi_{p_0}^{11}$ as the common Ansatz to all
these combinations.
\item[$\bullet$]Some combinations of three by
three Ans\"atze, taking as ground the previous combinations, where
one more Ansatz was included.
\item[$\bullet$]Beginning from that
combination of three Ans\"atze with the best total energy value,
new Ans\"atze were incorporated until there was a total of eleven
of them.
\end{itemize}

Each step in a particular minimization consists in keeping some
parameters as constant while others were varied. The election of
which parameters must be kept constant and which free was always
based on the physical interpretation of such parameters. In no way
the calculation is automatic, rather, a sophisticated artisan
treatment is required. With this procedure, a variational parameter
configuration was obtained in which the system´s total energy
value is minimum. Each minimization procedure for a given magnetic
field as a general statement took various months of calculations (see~\cite{Tesis}).

\section{Results }

In this section the total energy $E_T$, the binding
energy~\footnote{The binding energy is the necessary energy to
separate the system in its different components.} $E_b$ ($E_b
\equiv B - E_T$), and the equilibrium distance of protons
($R_3=R_1, R_4=R_2$) for the ground state $1\sigma_g$ and the
excited state $1\pi_u$ of the molecular ion $H_4^{3+}$ are
presented. The trial function used is shown in Eq.~(\ref{LFP}).
For a comparison with the results obtained
by Turbiner-L\'opez~\footnote{Although the first calculations for
this system are presented in ~\cite{LopezTur:2000}, the comparison
is made with the results not published by the same authors.}
the calculations were carried out considering two different conversion
factors to atomic units. These factors are
$B_0=2\textrm{.}35\times 10^9$\,G (the one adopted in this work)
and $B_0=2\textrm{.}3505\times10^9$\,G ( the one adopted in
~\cite{LopezTur:2000}).

\subsection{The Ground State $1\sigma_g$}

For the ground state $1\sigma_g$ each Ansatz presents a minimum in the energy curve for the finite parameter values $R_{1,\dots,4}$. In Table \ref{REBL} the results are present. The results carried out by Turbiner-L\'opez \cite{LopezTur:2000} (also present in Table \ref{REBL}) only consider the four terms that appear in Table \ref{TIF}.

\begin{table*}[thb]
\begin{center}
\begin{tabular}{| l  l  l  l  l  l  l |}
\hline
$B$(G)&$B_0(\times10^{9}$\,G)&$E_T$(Ry)&$E_b$(Ry)&$R_1$(a.u.)&$R_2$(a.u.)&\\
\hline \hline
$3\times 10^{13}$ & 2.35 & 12727.5422 & 38.4153 & 0.0635 & 0.2763 & Present \\
\hline \hline
                       & 2.35   & 18739.2526 & 43.7261 & 0.0574 & 0.2336 & Present\\
$4{.}414\times10^{13}$ & 2.3505 & 18735.2602 & 43.7230 & 0.0574 & 0.2336 & Present \\
                       &2.3505  & 18735.4675 & 43.5157 & 0.0572 & 0.2326 & Turbiner-L\'opez~\footnote{Not published.}\\
\hline
\end{tabular}
\end{center}
\caption{Results obtained for ground state $1\sigma_g$ of the
molecular ion $H_4^{3+}$ in the presence of a magnetic field $B$
for different values of the conversion factor $B_0$ and in
comparison with those obtained by Turbiner-L\'opez.} \label{REBL}
\end{table*}

\begin{figure}[thb]
\begin{center}
\begin{picture}(200,180)
\put(0,10){{\psfig{figure=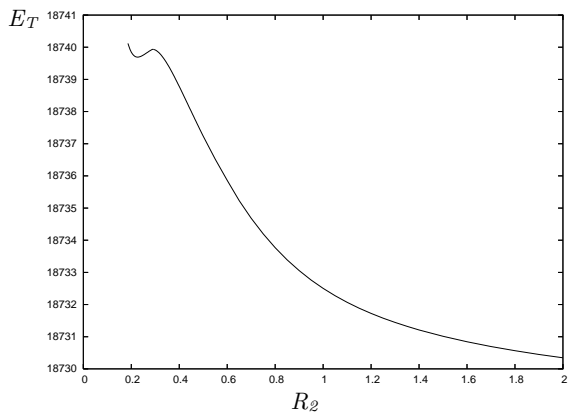,width=210pt,angle=-90}}}
\put(100,3){$ \mathit{\tiny R_2}$}
\put(-7,148){$\tiny \mathit{E_T}$}
\end{picture}
\end{center}
\vskip -20 pt \caption{Total energy curve of the molecular ion
$H_4^{3+}$ in the ground state $1\sigma_g$ as a function of
parameter R$_2$, which presents a minimum for finite equilibrium
distances. This in linear configuration and parallel to the
magnetic field $B=4\textrm{.}414\times10^{13}$~G.}
\label{La_curva}
\end{figure}

A total energy curve of the system as function of $R_2$ was obtained (see Fig.~\ref{La_curva}). For this the parameter $R_1$ is considered constant to the corresponding value of the minimum total energy configuration
(in symmetric configuration). The $R_3$ and $R_4$ parameters are now considered as variational parameters. In other words the symmetry restriction is eliminated. The curve presents a well defined minimum for finite internuclear distances in the
symmetric configuration. The height of the barrier is $\Delta E
\equiv E_{max}-E_{min} = 0\textrm{.}242$ Ry. The asymptotic
behavior of this curve for large $R_2$ shows that
$R_1+R_3 \thickapprox 0\textrm{.}1113$ (see Fig.~\ref{El_ion}) and
$R_4-R_3 \thickapprox 0\textrm{.}1110$. These values are very similar
to the equilibrium distances of the ion $H_3^{2+}$ (see Table \ref{T_Comparacion}). This indicates that a decay $H_4^{3+}\rightarrow H_3^{2+} + p$ is possible. The energy as function of $R_1$, $R_2$, $R_3$ and $R_4$,  
generates a surface. In particular the curve presented in
Fig.~\ref{La_curva} corresponds to the cut throughout the valley
of this surface.

\begin{figure}[thb]
\begin{center}
\begin{picture}(200,180)
\put(-15,-4){{\psfig{figure=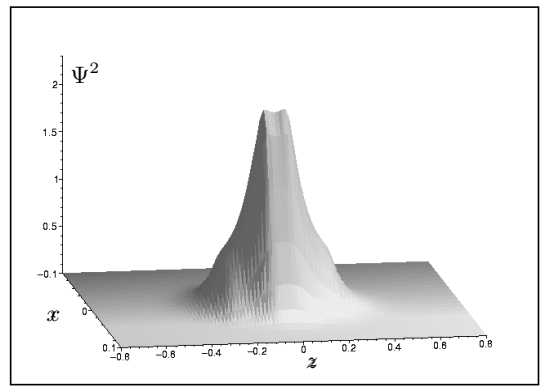,width=260pt,angle=-90}}}
\put(30,127){$\Psi^2$}
\put(21,37){$x$}
\put(119,19){$z$}
\end{picture}
\end{center}
\vskip -20 pt \caption{Electronic distribution of the molecular
ion $1\sigma_g$ in its ground state in a magnetic field $B =
4\textrm{.}414\times10^{13}$\,G, with $B_0 =
2\textrm{.}35\times10^{9}$\, G, in linear and parallel
configuration to the direction of the magnetic field, considering
that R$_1=$ R$_3$ and R$_2=$ R$_4$ (see Fig. 1).} \label{DEEB}
\end{figure}

The electronic distribution corresponding to ground state
$1\sigma_g$ is shown in Fig.~\ref{DEEB}. We can observed that
the electronic cloud is mainly concentrated around the
intermediate point to the protons.

\subsection{The Excited State  $1\pi_u$}

The calculations were made for the two values of the conversion factor $B_0$ and it is shown in Table \ref{REEL}. Also for this state, each Ansatz presents a minimum in the energy curve for the finite parameter values $R_{1,\dots,4}$. 

\begin{table*}[thb]
\begin{center}
\begin{tabular}{| l l l l l l |}
\hline
$B$(G) &$B_0(\times10^{9}$\,G) & E$_T$(Ry)\hspace{0.6cm} &  E$_b$(Ry)\hspace{0.8cm} &
R$_1$(a.u.)\hspace{0.2cm} & R$_2$(a.u.)\\
\hline \hline
$4{.}414\times10^{13}$ &2.35   & 18742.8185 & 40.1603  & 0.0651  & 5.5534 \\
 &2.3505 & 18738.8254 & 40.1578 & 0.0651  & 5.5537 \\
\hline
\end{tabular}
\end{center}
\caption{Results for the excited state $1\pi_u$ of the molecular
ion $H_4^{3+}$. For a magnetic field $B
=4\textrm{.}414\times10^{13}$\,G, with $B_0
=2\textrm{.}35\times10^{9}$\,G in linear, parallel and symmetric
configuration.} \label{REEL}
\end{table*}

\begin{figure*}[thb]
 \begin{center}
 \begin{picture}(225,500)
 \put(-20,351){\psfig{figure=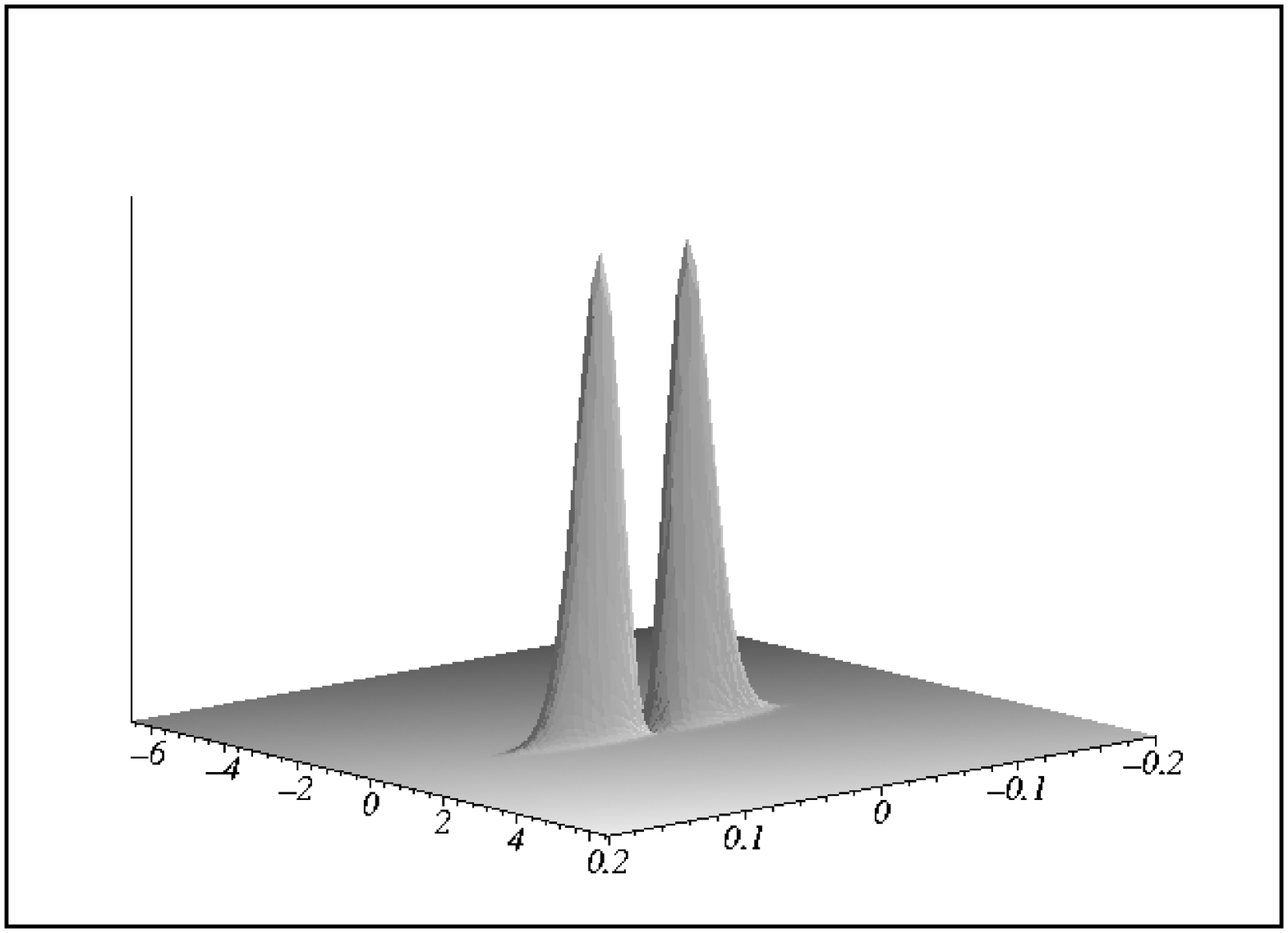,width=3.8in,angle=-90}}
 \put(28,480){$\Psi^2$}
 \put(50,375){$z$}
 \put(150,375){$x$}
 \put(195,505){$(a)$}
 \put(79,406){\large$\bullet$}
 \put(107,398){\large$\bullet$}
 \put(129,392){\large$\bullet$}
 \put(98,480){$1$}
 \put(115,483){$2$}
 \put(-24,-15){\psfig{figure=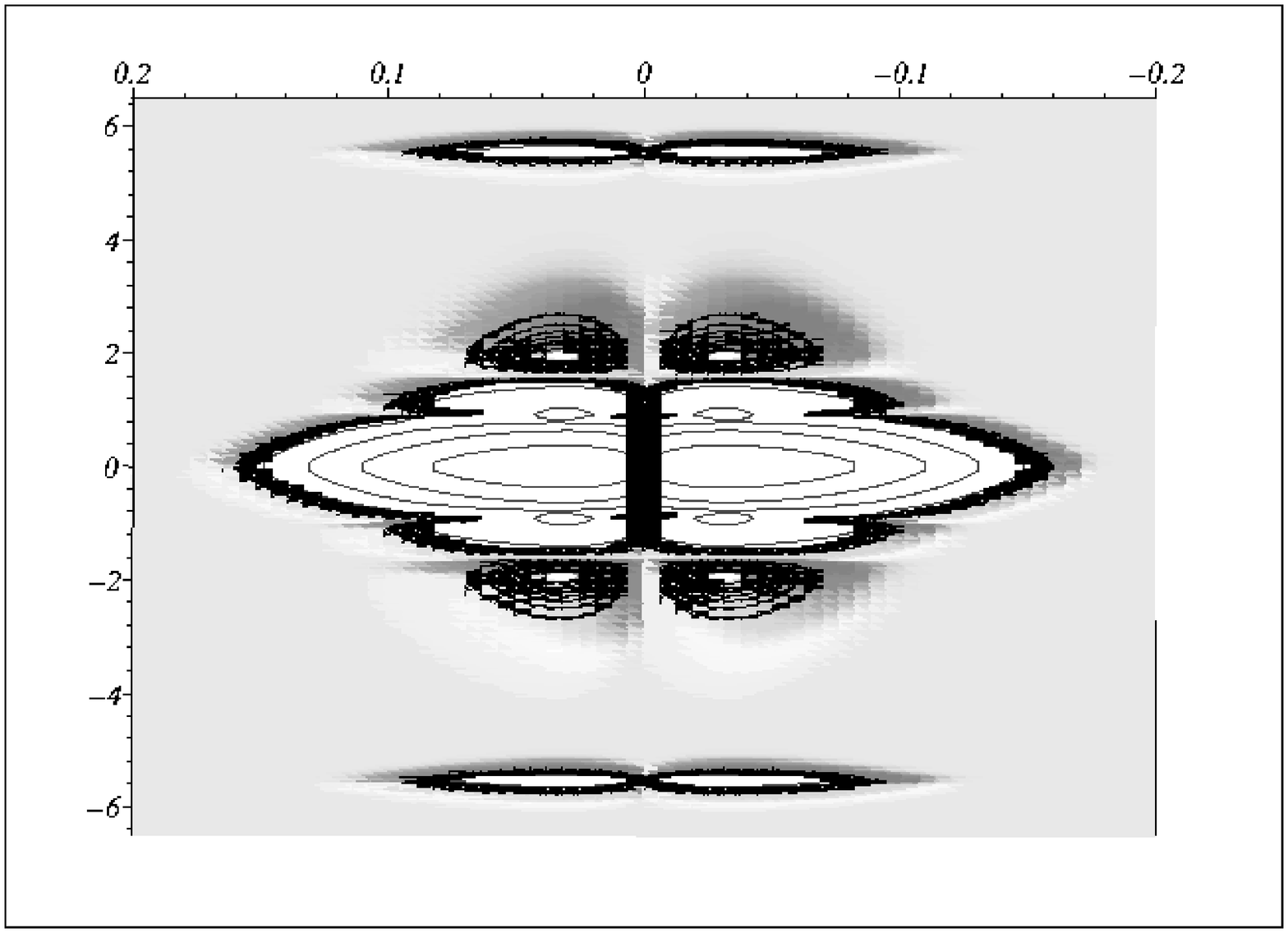,width=3.8in,height=5.3in,angle=-90}}
 \put(195,307){$(b)$}
 \put(105,310){$x$}
 \put(5,170){$z$}
 \put(88,169){$\blacklozenge$}
 \put(117,169){$\blacklozenge$}
 \put(90,172){\line(-2,1){55}}
 \put(120,172){\line(3,2){45}}
 \put(28,200){$1$}
 \put(170,205){$2$}
 \put(101,270){\LARGE$\bullet$}
 \put(101,167){\LARGE$\bullet$}
 \put(101,64){\LARGE$\bullet$}
 \end{picture}
 \end{center}
 \vskip -20 pt
 \caption{$(a)$ Electronic distribution of the molecular ion $H_4^{3+}$
for the excited state $1\pi_u$ in a magnetic field $B= 4.414
\times 10^{13}$\,G, with the conversion factor $B_0 = 2.35 \times
10^{9}$\,G. The points indicate the proton´s position (see text).
$(b)$ An enlargement of $10^{7}$ in vertical scale allows the
location of the other six pairs of peaks. The two most steep are
distinguished here with the numbers $1$ and $2$, and are indicated
with the symbol $\blacklozenge$. For a better visualization, a
perpendicular view to the plane $(x,z)$ is presented.
Normalization is not fixed.} 
\label{DEEX}
\end{figure*}

For the excited state $1\pi_u$ the electronic distribution is
shown in Fig.~\ref{DEEX}. The positions of the protons are
indicated by points in the plane $(x,z)$. The fact that apparently
there are only three points present and not four, it is due to the
fact that the distances between the internal protons is very small
in comparison with the whole dimension of the system (see
Table~\ref{REEL}). This electronic distribution presents seven
maximums (Fig.~\ref{DEEX}(b)) located in neighboring regions to
the protons. From these, those that surround internal protons (see
Fig.~\ref{El_ion}) are $\sim 10^{8}$ times more
pronounced that the minors of them. A comparison of the total
energy and the equilibrium distances~\footnote{This comparison
is carry out considering only the distance of the internal
protons for the system $H_4^{3+}$.} between molecular ions
$H_4^{3+}$ and $H_2^{+}$, both for the excited state
$1\pi_u$, shows that both are very similar
(see Table~\ref{T_Comparacion}), indicating  that $H_4^{3+} 
\rightarrow H_2^{+} + 2p$ could be a possible channel of decay of the
molecular system $H_4^{3+}$. If a comparison of the total
energy of the systems $H_4^{3+}$ and $H_3^{2+}$ is carry out (Table~\ref{T_Comparacion}), we found that the molecular ion $H_4^{3+}$
could have a possible channel of decay $H_4^{3+} \rightarrow H_3^{2+} + p$. Among the systems $H$, $H_2^{+}$, $H_3^{2+}$ and $H_4^{3+}$, the first of them, $H$, results as the one with the greatest total energy in the rank of the magnetic fields $0-4\textrm{.}414\times10^{13}$\,G, for which a possible decay $H_4^{3+} \rightarrow H + 3p$ is discarded.

\begin{table}[thb]
  \begin{center}
  \begin{tabular}{|l l l l l| }
  \hline
  \hspace{1cm}  & State\hspace{0.3cm} & E$_T$(Ry)\hspace{0.8cm} &
  E$_b$(Ry)\hspace{0.5cm} &  R$_{eq}$(a.u.) \\
  \hline \hline
  $H_2^+$ & $1\sigma_g $& 18728.477 & 54.5018& 0.1016 \\
          & $1\sigma_u $& 18750.07  & 32.912 & 2.021  \\
          & $1\pi_u    $& 18741.89  & 41.09  & 0.130  \\
          & $1\pi_g    $& 18757.273 & 25.7054& 2.237  \\
          & $1\delta_g $& 18747.527 & 35.407 & 0.148  \\
          & $1\delta_u $& 18761.18  & 21.80  & 2.230  \\
          & $2\sigma_g $& 18781.576 &  1.402 & 3.120  \\
  \hline \hline
  $H_3^{2+}$&$1\sigma_g$&18727.7475 &55.2312 &0.110   \\
          & $1\sigma_u $&Not exist &  &    \\
          & $1\pi_u    $&18742.7564 &40.2223 &0.145   \\
          & $1\pi_g    $&Not exist &  &    \\
          & $1\delta_g $&18748.9067 &34.0720 &0.167  \\
          & $1\delta_u $&Not exist &  &    \\
  \hline \hline
  $H_4^{3+}$&$1\sigma_g$& 18739.2526 &43.7261& 0.1149 \\
          &$1\pi_u$     & 18742.8185 &40.1603& 0.1302 \\
  \hline
  \end{tabular}
  \end{center}
  \caption{Comparison between molecular systems with one electron: $H_2^+$
  \cite{Turbiner:2004}, $H_3^{2+}$~\cite{Turbiner:2004c} and $H_4^{3+}$
(present) in different states. All of them in linear, symmetric
and parallel configuration in the presence of a magnetic field
\hbox{$B= 4\textrm{.}414\times10^{13}$}\,G
($B_0=2\textrm{.}35\times10^{9}$\,G). In $H_3^{2+}$ the
equilibrium distance R$_{eq}$ corresponds to the distance between
to adjacent protons. For the system $H_4^{3+}$ this was considered
as the separation between the two internal protons R$_{eq}=2$R$_1$
(see Fig.~\ref{El_ion}).}
  \label{T_Comparacion}
  \end{table}

Finally, it is important to mention that if we relax the  symmetry conditions ($R_{1,\dots,4}$ are arbitrary) in the minimization, the system returns to the symmetric configuration. This is fulfilled for the ground state ($m=0$) and for the excited state ($m=-1$). 


\section{Conclusions}

Applying the variational method, a detailed study of the  $1\sigma_{g}$ and $1\pi_{u}$ electronic states of the exotic molecular ion $H_4^{3+}$ in a strong magnetic field was carried out. The magnetic field $B$, was considered to be  constant and homogeneous.

\subsection{Ground State $1\sigma_g$}

An improvement in the variational energy of $\sim 0.2\%$ was obtained in comparison with the first calculations presents in the literature  (Turbiner-L\'opez). The equilibrium distances in both case are very similar. This in the symmetric configuration.
An important remark  is that the system $H_{4}^{3+}$ can already exist
for a magnetic field of intensity $B=3\times10^{13}$\,G.

\subsection{Excited State $1\pi_u$}

The existence of the excited state of the molecular ion $H_{4}^{3+}$ in a 
magnetic field $B=4\textrm{.}414\times10^{13}$\,G is seem possible.
This state presents a peculiarity in its equilibrium distances (see Table
\ref{REEL}). $R_2$ turns out to be $\sim 80$ times $R_1$.
The fact that the total energy of the molecular ions $H_4^{3+}$,
$H_3^{2+}$ and $H_2^{+}$ is very similar, indicates that the ion $H_{4}^{3+}$ is very unstable in this state. In this sense it is very unstable since any small disturbance could originate a decay. For which
the existence of other excited states does not seem possible, at least
for this magnetic field value.

{\bf Acknowledgments:}~The author wants to express his deep
gratitude to Dr. A.~Turbiner for the introduction to the subject
and the constant supervision during the investigations that led to
the Bachelor thesis which is based on the results of the present
work. I deeply indebted to Dr. J.C. L\'opez Vieyra for many
valuable discussions and a help in preparation of the manuscript.
As well I would like to thank Dr. N.~Guevara~L. for his help in
creating the computer code and his numerous comments during the
time when the work was done.

A financial support provided by the CONACYT through the project
No. 36650-E is gratefully acknowledged.

\end{document}